\documentclass[11pt]{article}
\usepackage[utf8]{inputenc}
\usepackage{geometry}
 \usepackage{graphicx}
 \usepackage{subfig}
  \usepackage{amsmath} 
  \graphicspath{{images33/}}
  \usepackage[usenames, dvipsnames]{color}
\usepackage{wrapfig}
\usepackage{boxedminipage}
\usepackage{enumerate,amsmath,amssymb,hhline,multirow,array}
\newcommand{\RNum}[1]{\uppercase\expandafter{\romannumeral #1\relax}}

\usepackage{romannum}
\usepackage{array}
\newcolumntype{M}[1]{>{\centering\arraybackslash}m{#1}}

\usepackage{fullpage}
\usepackage{xcolor}
\usepackage{amsmath}
\usepackage{amssymb}
\usepackage{graphicx}
\usepackage{mathrsfs}
\usepackage{wrapfig}
\usepackage{boxedminipage}
\usepackage{epsfig}
\usepackage{setspace}
\usepackage{float}
\usepackage{hyperref}
\usepackage{enumerate}
\usepackage[font=footnotesize,labelfont=rm]{caption}

\usepackage[numbers,sort&compress]{natbib}
\DeclareMathOperator*{\sech}{sech}

\def\be{\begin{equation}}
\def\ee{\end{equation}}
\def\bea{\begin{eqnarray}}
\def\eea{\end{eqnarray}}

\numberwithin{equation}{section}

 \newcommand{\RN}[1]{%
   \textup{\uppercase\expandafter{\romannumeral#1}}%
 }
\begin{document}

\pagenumbering{arabic}
%{\tt hep-th/yymmnnn}

\vskip 2cm

\begin{center}
{\Large \bf Chaos in Charged Gauss-Bonnet AdS Black Holes in Extended Phase Space }
\end{center}

\vskip .2cm

\vskip 1.2cm

%\centerline
\begin{center}
	{ {\bf	Sandip Mahish} \footnote{ sm19@iitbbs.ac.in} and {\bf Chandrasekhar Bhamidipati}\footnote{ chandrasekhar@iitbbs.ac.in} 
	\\ 
	\vskip 0.7cm
	School of Basic Sciences\\
	}
{
Indian Institute of Technology Bhubaneswar \\ Jatni, Khurda, Odisha, 752050, India}
\end{center}

\vskip 1.2cm
\vskip 1.2cm
\centerline{\bf Abstract}
\noindent

We study the onset of chaos due to temporal and spatially periodic perturbations in charged Gauss-Bonnet AdS black holes in extended thermodynamic phase space, by analyzing the zeros of the appropriate Melnikov functions. Temporal perturbations coming from a thermal quench in the unstable spinodal region of P-V diagram, may lead to chaos, when a certain perturbation parameter $\gamma$ saturates a critical value, involving the Gauss-Bonnet coupling $\alpha$ and the black hole charge $Q$. A general condition following from the equation of state is found, which can rule out the existence of chaos in any black hole. Using this condition, we find that the presence of charge is necessary for chaos under temporal perturbations. In particular, chaos is absent in neutral Gauss-Bonnet and Lovelock black holes in general dimensions. Chaotic behavior continues to exist under spatial perturbations, irrespective of whether the black hole carries charge or not.

%\begin{quote}
%\noindent
%\end{quote}
\newpage
\setcounter{footnote}{0}
\noindent

\baselineskip 15pt

\section{Introduction}

\noindent
Black hole solutions and their thermodynamics in General Relativity have thrown up remarkable surprises and continue to be an intriguing area of research. In particular, phase transitions of black holes in a variety of backgrounds, such as Anti de Sitter (AdS) space-time have been actively pursued, purely from gravity point of view and also with holographic motivations in mind~\cite{Bekenstein:1973ur}-\cite{Gubser:1998bc}. More recently, treating the cosmological constant as a dynamical thermodynamical variable (pressure), an extended phase thermodynamics has been proposed, where the first law of black hole mechanics gets modified by a new $pdV$ term~\cite{Kastor:2009wy}-\cite{Johnson:2018amj}. Study of PV critical behavior of various black holes confirms the existence of an exact map of black hole small/large phase transitions to the Van der Waals liquid/gas system~\cite{Chamblin}-\cite{Gunasekaran:2012dq}. \\

\noindent
It is known that chaos is unavoidable in certain dynamical systems in nature, including black hole physics and cosmology~\cite{Bombellitf}-\cite{Chen:2016tmr}. There have been several past works probing chaotic behavior in black holes by various methods, such as, computation of Lyapunov exponents to study stability of orbits, quasinormal modes in Reissner-Nordstrom and Gauss-Bonnet black holes~\cite{Cardoso:2008bp,Konoplya:2017wot}, and Melnikov's~\cite{Melnikov-Holmes} method in the context of geodesic motion~\cite{Bombellitf,Letelier,Manuele}. However, the study of chaos in the context of black hole thermodynamics and phase transitions has only just started emerging~\cite{Chabab:2018lzf}, partly due to the recent developments where a pressure term in the first law is included, making the connection with Van der Waals system exact\cite{Chamblin}-\cite{Gunasekaran:2012dq}. In~\cite{Chabab:2018lzf}, the Melnikov method used in dynamical systems~\cite{Melnikov-Holmes}-\cite{Holmes1990}, developed in the context of Van der Waals system~\cite{Slemrod}, was applied to the case of black holes in extended phase space to extract useful information about the presence of chaos.  Temporal and spatial period perturbations were introduced in the PV thermodynamic phase space and the presence of chaos was detected from the study of zeros of Melnikov function.  A bound involving charge of the black hole was also found, beyond which the system becomes chaotic.  \\

\noindent
In this letter, we take these issues forward by studying chaos in extended thermodynamic phase of black holes, after incorporating the effects of higher curvature terms in Einstein Action. We focus on the case of Guss-Bonnet(GB) black holes, but the results are also spelled out for Lovelock black holes. Gauss-Bonnet and Lovelock terms are quite important in various contexts such as, semi-classical quantum gravity, low energy effective action of string theory and next to leading order large N corrections of boundary conformal field theory (CFT) studies in holography~\cite{Lovelock:1971yv}-\cite{Mo:2014qsa}. They are known to have given interesting insights in to the corrections to black hole entropy, viscosity to entropy ratio and several other recent developments in extended phase space~\cite{Sen:2005iz,Brigante:2007nu,Cai:2009zv,Nojiri:2001aj,Johnson:2015ekr,Bhamidipati:2017nau}.  Chaotic dynamics of test objects and instability of certain orbits, in particular, in the context of holography and Gauss-Bonnet theories has also been explored before~\cite{Konoplya:2017wot,Kim:2013xu,Ma:2014aha}, however, not from thermodynamic point of view.  The extended phase thermodynamics of Gauss-Bonnet black holes in AdS (where the cosmological constant is taken to be dynamical) and its connection to the Van der Waals liquid/gas system via PV criticality is now well studied~\cite{Cai:2013qga}. Following the study of chaotic behavior for Reissner-Nordstrom black holes in AdS, it is important to know whether the behavior found in~\cite{Chabab:2018lzf} is a generic feature of systems exhibiting Van der Waals type phase transitions. With this motivation,  we thus study chaotic dynamics in Gauss-Bonnet and other higher derivative theories of gravity, with the inclusion of an additional parameter, such as the Gauss-Bonnet coupling, in addition to charge (considered in~\cite{Chabab:2018lzf}), and find that there appears a new inequality which governs the existence of chaos. We also show that neutral Gauss-Bonnet black holes in five and higher dimensions, in contrast, do not show chaotic behavior under temporal perturbations, despite the fact that Van der Waals type phase transition and PV criticality exists~\cite{Cai:2013qga}. We generalize this result to generic black holes systems which have an extended thermodynamic phase space description and starting from the equation of state, we find a new relation which can be used to rule out chaos. However, chaotic behavior under spatial perturbations in the unstable thermodynamic region, continues to exist for charged, as well as, neutral black holes. The results are also extended to Lovelock black holes in various dimensions.\\

\noindent
Rest of the paper is organized as follows. In section-\ref{melgb}, we recall the definition of Melnikov function and  few known aspects of thermodynamics of GB black holes in extended phase space formalism. Section-\ref{temp} deals with the effect of having a small temporal perturbation in the spinodal region of GB black hole thermodynamic phase space. We first obtain the analogue Hamiltonian system starting from the equation of state of the GB black hole, leading to the determination of homoclinic/heteroclinic orbits. Using the solutions for these orbits, the Melnikov function is computed explicitly and its zeros are analyzed, which give a bound on the parameter $\gamma$ (following from a small temporal perturbation, to be introduced in section-\ref{temp}) for existence of chaotic behavior. This bound is also discussed for Lovelock black holes in higher dimensions and a general condition for ruling out chaos in any black hole is obtained. In section-\ref{spac}, the effect of a small spatial perturbation leading to the onset of chaos is discussed for GB black holes. Section-\ref{conclusions} contains conclusions.

%%%%%%%%%%%%%%%%%%
\section{Melnikov's Method and Charged Black Holes in AdS}\label{melgb}
We start in subsection-(\ref{mel}) by summarising the basic technique due to Melnikov for studying the onset of chatotic behavior in Hamiltonian systems. In the following subsection-(\ref{gbbh}), we collect main results on charged Gauss-Bonnet black holes in AdS. These are then used for the computation of Melnikov function in sections-(\ref{temp}) and (\ref{spac}), for studying temporal and spatial perturbations, respectively in charged Gauss-Bonnet black holes.

%%%%%%%%%%%%%%%%%%%
\subsection{Melnikov's Method for Perturbation of Hamiltonian Systems}\label{mel}
\begin{figure}
	\begin{center}
		{\centering
			\subfloat[]{\includegraphics[width=4cm,height=3cm]{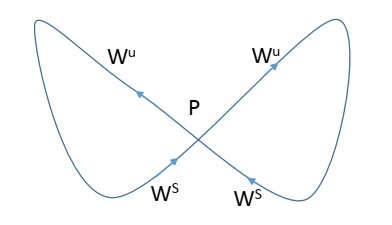} } \hspace {1.5cm}
			\subfloat[]{\includegraphics[width=4cm,height=3cm]{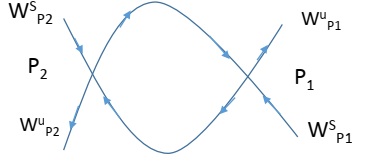} } 
			\caption{(a)Homoclinic orbit . (b) Heteroclinic orbit }  \label{heterohomo}
		}
	\end{center}	
\end{figure}

%%%%%%%%%%%%%%%%%%%%
To understand the Melnikov method, it is useful to start from an evolution equation for a displacement function $x(t)$ as follows:
   \begin{equation}\label{evolution}
  \dot{x}=f_0(x)+\epsilon f_1(x,t), \quad  x \in \Re^{2n}  \, ,
 \end{equation}
with the following assumptions. First, $\epsilon \ll 1$, corresponding to a small perturbation and the function $ f_1(x,t)$, is taken to be periodic in $t$. Second, the unperturbed system is Hamiltonian with smooth flow, conserving energy and contains a fixed point which is a homoclinic orbit\footnote{The Melnikov method also works for heteroclinic orbits connecting two saddle points and irrespective of whether the solution $x_0(t)$ is known analytically or not.}. There are further non-resonance assumptions on the function $f_1(x,t)$, which are necessary for smooth period perturbations and are given in the appropriate sections below.  Figure-(\ref{heterohomo}) shows sample plots of homoclinic and heteroclinic orbits. For homoclinic orbits, the stable $W^s$ and unstable $W^u$ manifolds of the saddle connect to each other at the hyperbolic fixed point $P$, while for heteroclinic orbits, the stable manifold of one saddle joins the unstable manifold of the other saddle, as seen in figure-(\ref{heterohomo}).
	\begin{figure}[h]
	\begin{center}
		{\centering
			\subfloat[]{\includegraphics[width=3cm,height=3cm]{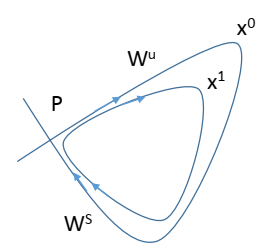} } \hspace {1.5cm}
			\subfloat[]{\includegraphics[width=3cm,height=3cm]{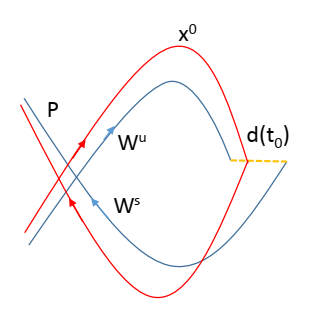} } 
			\caption{Homoclinic orbit: (a) Before perturbation. (b) After perturbation }  \label{homoperturbed}
		}
	\end{center}	
		\end{figure}
Now, under a temporal/spatial perturbation of the system,  there will be infinite number of complicated intersection points of stable and unstable manifolds (as it takes an infinite time to approach a saddle point). If the unperturbed homoclinic orbit is considered to be a curve parametrized by time, then at time $t_{0}$, the stable and unstable manifolds are seperated by a perpendicular distance (as shown in figure-(\ref{homoperturbed})) given as 
$$d(t_{0})=\frac{\epsilon M(t_{0})}{|\textbf{f}(x_{0}(0))|}.$$ Here,  $M(t_{0})$ is known to be the the Melnikov function and its explicit form can be shown to be~\cite{Melnikov-Holmes}-\cite{Holmes1990}:
 \begin{equation} \label{melnikov}
 M(t_{0})=\int_{-\infty}^{+\infty} \, \text{f}_0^{T}\left( x_{0}(t-t_{0})\right)  \Omega_n \text{f}_1\left( x_{0}(t-t_{0}),t\right)dt,
 \end{equation}\\
 with,
 \begin{equation}
\Omega_{n=2}=\begin{pmatrix} 0 & 1 & 0 & 0 \\-1 & 0 & 0 & 0\\0 & 0 & 0 & 1 \\ 0 & 0 & -1 & 0 	\end{pmatrix}
\quad \text{and} \quad
\Omega_{n=1}=\begin{pmatrix} 0 & 1  \\-1 & 0
	\end{pmatrix} \, .
 \end{equation}
Here, the subscript $1$ and $2$ stand for the number of degrees of freedom appearing in temporal and spatial perturbations, respectively. Melnikov function $M(t_{0}) $, is thus an estimate of the distance $d(t_{0})$ for the transverse intersections of stable and unstable orbits. If $ M(t_{0})$ has a simple zero as a function of $t_0$, then for $\epsilon >0$ and for suitably small value, the stable and unstable manifold of the Hamiltonian system intersect transversally~\cite{Melnikov-Holmes}-\cite{Holmes1990}. From Smale-Birkhoff theorem~\cite{sb}, presence of such intersecting orbits implies that the Poincare Map has a invariant hyperbolic set: a Smale horseshoe, which is an indicator of chaos~\cite{Holmes1990,Smale}.

%%%%%%%%%%%%%%%%%%%
\subsection{Charged Gauss-Bonnet Black Holes in AdS}\label{gbbh}
\begin{figure}[h]
 	\centering
 	\includegraphics[width=7cm,height=5cm]{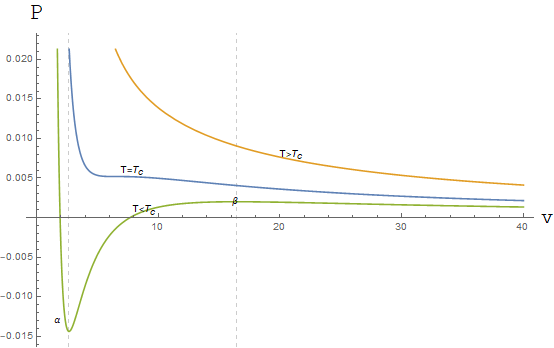}
 	\caption{P-v diagram for the Gauss-Bonnet AdS Black hole }
 	\label{fig1}
 \end{figure}

%%%%%%%%%%%%%%%%%
\noindent
We start with some preliminaries on thermodynamics of black holes in extended phase space and defining the spinodal region where chaos is found. The Einstein-Maxwell action with a Gauss Bonnet term and a cosmological constant $\Lambda$,  in $d$ dimensions is as follows:
\begin{equation}
S=\frac{1}{16\pi}\int d^{d}x\sqrt{-g}[R-2\Lambda+\alpha_{GB}(R_{\mu\nu\gamma\delta}R^{\mu\nu\gamma\delta}-4R_{\mu\nu}R^{\mu\nu}+R^{2})-4\pi F_{\mu\nu}F^{\mu\nu}],
\label{gauss bonnet action}
\end{equation}
where $\alpha_{GB}$ is Gauss Bonnet coupling and 
%which has dimension of $[length]^{2}$, 
$\Lambda=-\frac{(d-1)(d-2)}{2l^{2}}$. $F_{\mu\nu}$ is the Maxwell field strength, defined as $F_{\mu\nu}=\partial_{\mu} A_{\nu}-\partial_{\nu} A_{\mu}$, with the vector potential $A_{\mu}$. Here we will mostly consider the case with $\alpha_{GB}\geq 0$. The Gauss Bonnet term, proportional to $\alpha_{GB}$ in the above action, is a topological term in four dimensions and hence we take $d\geq5$. The solution for a static charged GB black hole is given as:
\begin{equation}
ds^{2}=-f(r)dt^{2}+\frac{dr^{2}}{f(r)}+r^{2}d\Omega_{d-2}^{2}\, ,
\end{equation}
where $d\Omega_{d-2}^{2}$ is a line element of $(d-2)$ dimensional maximally symmetric Einstein space with volume $\Sigma_{k}$ where k can be 1,0,-1, corresponding to spherical, Ricci flat and hyperbolic topology of black hole horizon, respectively. We will mainly deal with horizon of spherical topology. The general metric function is given by \cite{Cai:2013qga}
\begin{equation}
f(r)=k+\frac{r^{2}}{2\alpha}\Big(1-\sqrt{1+\frac{64\pi\alpha M}{(d-2)\Sigma_{k}r^{d-1}}-\frac{2\alpha Q^{2}}{(d-2)(d-3)r^{2d-4}}-\frac{64\pi \alpha P}{(d-1)(d-2)}}\Big) \, .
\end{equation}
Here $\alpha=(d-3)(d-4)\alpha_{GB}$; $M$ and $Q$ are mass and charge of black hole, and the pressure $P=-\frac{\Lambda}{8\pi}$. Notice that we have considered the cosmological constant to be a thermodynamic variable and replaced it with pressure as is the norm in extended thermodynamic phase space approach. The equation of state can be written as~\cite{Cai:2013qga}:
\begin{equation}
P=\frac{d-2}{4r}(1+\frac{2k{\alpha}}{r^2})T-\frac{(d-2)(d-3)k}{16\pi r^2}-\frac{(d-2)(d-5)k^2{\alpha}}{16\pi r^4}+\frac{Q^2}{8\pi r^{2d-4}}.
\label{eqnstateGBQ}
\end{equation}
%Form now on we will deal with charged Gauss Bonnet black hole in 5 dimension,hence $d=5$. For that case $\Sigma_{k}=2\pi^{2}$.Since for 5 dimension the metric function will be $$f(r)=k+\frac{r^{2}}{2\alpha}\Big(1-\sqrt{1+\frac{32\alpha M}{3\pi r^{4}}-\frac{\alpha Q^{2}}{3r^{6}}-\frac{16\pi\alpha p}{3}}\Big)$$ 
Thus, the first law in extended phase space is:
\begin{equation}
dM=TdS+\Phi dQ+VdP+\mathcal{A}d\alpha \, ,
\end{equation}
where $S$ is the entropy, $\Phi$ is the electric potential and $\mathcal{A}$ is conjugate to the GB coupling $\alpha$.  Hawking temperature $T$ the thermodynamic volume $V$ are given respectively as 
\begin{equation}
T=\frac1{4\pi}f'(r)=\frac{16\pi P r ^4/3+2k r ^2-\frac{2 Q^2}{3r ^{2}}}{4\pi r  (r ^2+2k\alpha)}\, ,
\label{hawking temperature}
\end{equation}
and
\begin{equation}
V=\frac{\Sigma_k r_h ^{d-1}}{d-1} \, .
%=\frac{2\pi^{2}r^{4}}{4}\, .
\label{Thermo vol}
\end{equation}
Equation of state in five dimensions is thus:
\begin{equation}
P=\frac{T}{v} \left(1+\frac{32 \alpha  k}{9 v^2}\right)-\frac{2 k}{3 \pi  v^2}+\frac{512 Q^2}{729 \pi
	v^6}\, ,
\label{eqnstate}
\end{equation}
where the specific volume $v=\frac{4r}{d-2}=\frac{4r}{3}$.
The Melnikov method is well suited for studying chaotic behavior in the black hole systems which follow Van der Waals equation for phase transition in the extended phase space formalism. 
\noindent
To study the behavior of the system in spinodal region, the $P-v$ phase diagram is introduced in figure-(\ref{fig1}), where the labeling of different points is explained below.
Denoting $\delta P = \partial P(v,T_{0})/\partial v$ and for a temperature $T^{\prime}$ below critical temperature, the  phase space of specific volume, i.e., $v\in[0,\infty]$),  is divided in to three  regimes.  $\left[0,\alpha \right]$ corresponds to the region where small black holes exist, i.e., the fluid being in liquid phase, i.e., $\delta P< 0$. $\left[\alpha,\beta\right]$ corresponds to an unstable region, where small and large black hole phase co-exists, i.e., $\delta P> 0$. The two points $\alpha$ and $\beta$ are determined by $\delta P\Big{|}_{v=\alpha}=\delta P\Big{|}_{v=\beta}=0$.  This is the main region of interest in the present case, called the spinodal domain, where a temporal or spatial periodic perturbation leads to chaos under certain conditions, to be discussed below. $\left[\beta,\infty\right]$ corresponds to the large black hole domain, i.e., the fluid being vapor and where $\delta P <0$.

%%%%%%%%%%%%%%%%%%%%%%%%%%

\section{Temporal Chaos in Spinodal region} \label{temp}

%%%%%%%%%%%%%%%%%%%%%%%%%%%%

\begin{figure}[h]
  	% \begin{wrapfigure}{l}{0.3\textwidth}
  	\begin{center}
  		{\centering
  		\subfloat[]{\includegraphics[width=1.5in]{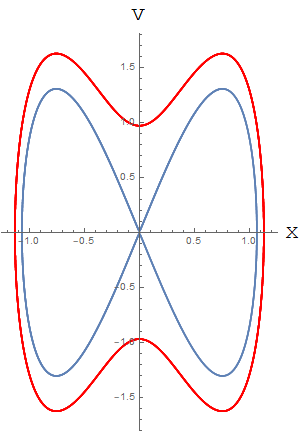} } \hspace {1.0cm}
  			\subfloat[]{\includegraphics[width=1.4in]{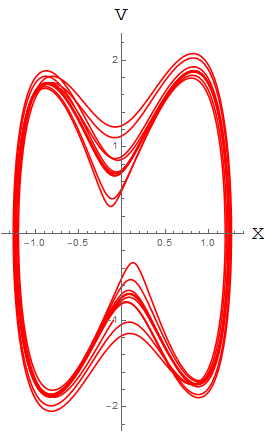} } \hspace {1.0cm}
  			\subfloat[]{\includegraphics[width=1.3in]{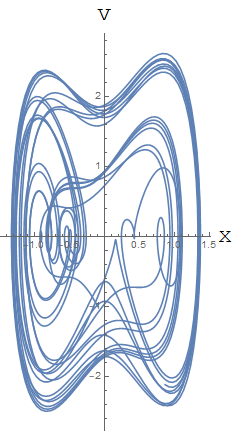} }
  			\caption{Plots show 
time evolution in the phase space of velocity vs displacement for (a) $\epsilon=0$ (b) $\epsilon \neq 0$,$\gamma < \gamma_{\rm critical}$ (c) $\epsilon \neq 0$,  $\gamma > \gamma_{\rm critical}$}   \label{fig:gammacritical1}
  		}
  	\end{center}
  	%\end{wrapfigure}
  \end{figure}

\begin{figure}[h]
  	% \begin{wrapfigure}{l}{0.3\textwidth}
  	\begin{center}
  		{\centering
  		\subfloat[]{\includegraphics[width=2.3in]{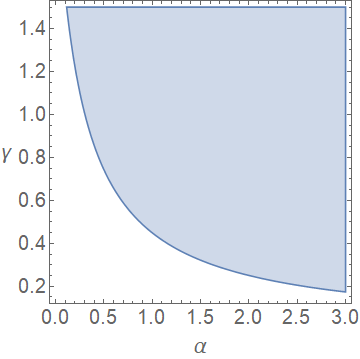} } \hspace {1.5cm}
  			\subfloat[]{\includegraphics[width=2.2in]{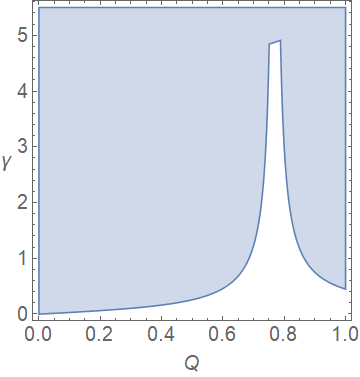} } 
  			\caption{Shaded region denotes onset of chaotic motion: $ v_0=2.75, T_0=0.01$ (a) $\gamma$ vs $\alpha$ plot for charged GB black holes in $d=5$, for $Q=1$ (b)$\gamma$ vs $Q$ plot for  charged GB black holes in $d=5$ for $\alpha= 1$  }   \label{fig:gammacritical2}
  		}
  	\end{center}
  	%\end{wrapfigure}
  \end{figure}

Here, we study the effect of a small temporally periodic perturbation, when the system is quenched to the unstable spinodal region. We first compute the Hamiltonian for the fluid flow using the black hole equation of state and obtain the Melnikov function, which contains information about the onset of chaos. Let us start by considering a specific volume $v_{0}$ corresponding to an isotherm $T_{0}$, which is fluctuated as follows~\cite{Slemrod,Chabab:2018lzf}:
\bea \label{pert}
&& T=T_{0}+\epsilon \, \gamma \cos(\omega t)\cos(X) ~~~{\rm with}~~~\epsilon<<1 \, . 
\eea
The fluid flow is assumed to be taking place along the x-axis in a tube of unit cross section with fixed volume, which contains a total of mass $2\pi/q$ of fluid in a volume $(2\pi/q)v_0$, where $q>0$ is a constant~\cite{Slemrod}. The fluid is further assumed to be thermoelastic, slightly vicious and isotropic with an additional stress following from the van der Waals-Korteweg theory of capillarity~\cite{Slemrod}.
Here, $X$ represents a column of black hole of unit cross section taken between certain points and the details together with other assumptions are similar to earlier considerations given in~\cite{Slemrod,Chabab:2018lzf}. In present case the Hamiltonian is symbolically given to be~\cite{Slemrod,Chabab:2018lzf}:
\begin{equation}
H=\frac{1}{\pi}\int_{0}^{2\pi}\Big[\frac{u^{2}}{2}+\mathcal{F}(v,T)+\frac{Aq^{2}}{2}\Big(\frac{\partial v}{\partial X}\Big)^{2}\Big]dX
\label{Hamiltonian}
\end{equation}
where $u(x,t)=x_{t}(X,t)$ is the velocity of the reference fluid particle and $A$ is a constant. Here, 
\begin{align}
\mathcal{F}(v,T)&=-\int \bar{P}(v,T)\, dv \,,
\label{free energy}
\end{align}
with 
\bea \label{EffectiveEOS}
\bar{P}(v,T) =  P(v,T)\frac{dV}{dv} = \frac{4 \pi Q^2}{9 v^3} + 
\frac{9}{128} \pi^2 (-6 k v + 9 T v^2 + 32 k T \alpha))\, .
\eea
%\begin{align}
%\mathcal{F}(v,T) =-\int P(v,T) dv \,
%\label{free energy}
%\end{align}
$\bar{p}(v,T)$ is an effective equation of state obtained by replacing $v$ in terms of the thermodynamic volume $V=\frac{\pi v^3}{6}$ before performing the integral. Let us also note that the use of effective equaion of state in eqn. (\ref{EffectiveEOS}) is important for existence or non-existence of chaos below. This is important as the Gibbs free energy written in terms of thermodynamic volume remains unchanged during phase transition and it is the combination $PdV$ that has the right scaling~\cite{Lan:2015bia,Xu:2015hba}. 
Following the approach in~\cite{Slemrod,Chabab:2018lzf}, ignoring coefficients of order ${\mathcal O}(1/v^4)$ in Taylor series expansion, the Hamiltonian can be computed to be:
\begin{align}\label{HamG}
H(x,u)&=\frac{u_{1}^{2}+u_{2}^{2}}{2}-\frac{\bar{P}_{v}}{2}(v_{0},T_{0})(x_{1}^{2}+x_{2}^{2})-\frac{\bar{P}_{v,v}}{4}(v_{0},T_{0})x_{1}^{2}x_{2}\nonumber\\&-\frac{\bar{P}_{v,v,v}}{32}(v_{0},T_{0})(x_{1}^{4}+x_{2}^{4}+4x_{1}^{2}x_{2}^{2})-\bar{P}_{T}(v_{0},T_{0})\epsilon\gamma \cos(\omega t)x_{1}\nonumber\\
&-\bar{P}_{v,T}(v_{0},T_{0})\epsilon\gamma \cos(\omega t)x_{1}x_{2}-\frac{\bar{P}_{v,v,T}}{24}(v_{0},T_{0})3\epsilon\gamma \cos(\omega t)x_{1}(x_{1}^{2}+2x_{2}^{2})\nonumber\\&+\frac{Aq^{2}}{2}(x_{1}^{2}+4x_{2}^{2}) \, .
\end{align}   
Let us note that the above form of the Hamiltonian is quite generic and is valid for any black hole in extended phase space thermodynamics. For the particular case of charged GB black holes, the relevant
Hamiltonian is found to be:
\begin{align}\label{Ham}
H(x,u)&=\frac{u_{1}^{2}+u_{2}^{2}}{2}+\frac{Aq^{2}}{2}(x_{1}^{2}+4x_{2}^{2})-\frac{9\pi^2}{128} \epsilon\gamma \cos(\omega t)(32k\alpha+9v_{0}^{2})x_{1}-\frac{81\pi^2}{128} \epsilon\gamma \cos(\omega t)v_{0}x_{1}x_{2}\nonumber\\&-\frac{1}{4}(\frac{81\pi^{2}T_{0}}{64}+\frac{16\pi Q^{2}}{3v_{0}^{5}})x_{1}^{2}x_{2}-\frac{1}{2}(-\frac{4\pi Q^{2}}{3v_{0}^{4}}+\frac{9\pi}{128}(-6k+18\pi T_{0}v_{0}))(x_{1}^{2}+x_{2}^{2})\nonumber\\&-\frac{81\pi^{2}}{512}\epsilon\gamma\cos(\omega t)x_{1}(x_{1}+2x_{2}^{2})+\frac{5\pi Q^{2}}{6v_{0}^{6}}(x_{1}^{4}+x_{4}+4x_{1}^{2}x_{2}^{2}) \, .
\end{align} 
Here, $(x_{1},x_{2})$ and $(u_{1},u_{2})$ are position and velocities of first two modes, and the corresponding equation of motion are:
\bea\label{eom1}
\begin{aligned}
\dot{x}_{1} &=\frac{\partial H_{2}}{\partial u_{1}}=u_{1} \, ,\\
\dot{x}_{2} &=\frac{\partial H_{2}}{\partial u_{2}}=u_{2} \, ,
\end{aligned}
\eea
\bea\label{eom2}
\begin{aligned}
\dot{u}_{1} &=-\frac{\partial H_{2}}{\partial x_{1}}-\epsilon\mu_{0}qu_{1} 
= \frac{9\pi^{2}}{128}\epsilon\gamma\cos(\omega t)(32k\alpha+9v_{0}^{2})-Aq^{2}x_{1}\\
&+(-\frac{4\pi Q^{2}}{3v_{0}^{4}}+\frac{9\pi^{2}}{128}(-6k+18T_{0}v_{0}))x_{1}+\frac{81\pi^{2}}{256}\epsilon\gamma\cos(\omega t)x_{1}^{2}\\
&+\frac{81\pi^{2}}{128}\epsilon\gamma\cos(\omega t)v_{0}x_{2}+\frac{1}{2}(\frac{81\pi^{2}T_{0}}{64}+\frac{16\pi Q^{2}}{3v_{0}^{5}})x_{1}x_{2}+\frac{81\pi^{2}}{512}\epsilon\gamma\cos(\omega t)(x_{1}^{2}+2x_{2}^{2})\\
&-\frac{5\pi Q^{2}}{6v_{0}^{6}}(4x_{1}^{3}+8x_{1}x_{2}^{2})-q\epsilon\mu_{0}u_{1}\, ,
\end{aligned}
\eea
\bea
\begin{aligned} \label{eom4}
\dot{u}_{2} =-\frac{\partial H_{2}}{\partial x_{2}}-4\epsilon\mu_{0}qu_{2} &=+\frac{81}{128}\pi^{2}\epsilon\gamma\cos(\omega t)v_{0}x_{1}+\frac{1}{4}(\frac{81\pi^{2}T_{0}}{64}+\frac{16\pi Q^{2}}{3v_{0}^{5}})x_{1}^{2}-4Aq^{2}x_{2}\\
&+(-\frac{4\pi Q^{2}}{3v_{0}^{4}}+\frac{9\pi^{2}}{128}(-6k+18T_{0}v_{0}))x_{2}+\frac{81\pi^{2}}{128}\epsilon\gamma\cos(\omega t)x_{1}x_{2}\\
&-\frac{5\pi Q^{2}}{6v_{0}^{6}}(4x_{2}^{3}+8x_{1}^{2}x_{2})-4q\epsilon \mu_{0}u_{2} \, .
\end{aligned}
%\right.
\label{eq21}
\eea
Writing $z=(x_{1},x_{2},u_{1},u_{2})^{T}$, eqns. (\ref{eom1})-(\ref{eq21}) can be written in a compact form as
$\dot{z}(t)=f_{0}(z)+\epsilon f_{1}(z,t) $;
$f_{1}$ is periodic in t, and the unperturbed system with $\epsilon=0$ is given as $\dot{z}(t)=f_{0}(z)$. If we linearize the unperturbed system about $z=0$, we get $\dot{z}_{L}(t)=Lz_{L}(t)$.  The matrix $L$ can be computed to be~\cite{Slemrod,Chabab:2018lzf}:
\begin{equation}
\textbf{L}=\begin{pmatrix} 0 & 0 &  1 & 0 \\0 & 0 & 0 & 1\\-A q^2+\psi& 0 & -\epsilon\mu_{0}q & 0 \\ 0 & -4 A q^2+\psi& 0 & -4\epsilon\mu_{0}q
\end{pmatrix},
\end{equation} 
with eigen values:
\begin{eqnarray}\nonumber
\lambda_{1,2}&=&\frac{-\epsilon\mu_{0}q}{2}\pm\frac{1}{2}[\epsilon^{2}\mu_{0}^{2}q^{2}-4(A q^2-\psi)]^{\frac{1}{2}},\\ \nonumber
\lambda_{3,4}&=&-\frac{4\epsilon\mu_{0}q}{2}\pm[4\epsilon^{2}\mu_{0}^{2}q^{2}-(4A q^2-\psi)]^{\frac{1}{2}} \, .
\end{eqnarray}
%=\bar{P}_{v}\Big{|}_{v_{0},T_{0}}$$
Here, $$\psi=-\frac{4 \pi Q^2}{3 v_{0}^4} + \frac{9 \pi}{128} (-6 k + 18\pi T_{0} v_{0}) \, .$$
Stability of the nodes depends on $q^{2}$. For $q^{2}<\frac{\psi}{A}$, one notes that $\lambda_{1}>0$, $\lambda_{2}<0$ and both are real; while for $q^{2}>\frac{\psi}{4A}$, both the eigen values $\lambda_{3,4}=-\frac{4\epsilon\mu_{0}q}{2}\pm\iota[(4A q^2-(\psi)]^{\frac{1}{2}}$ are imaginary. Regarding $\lambda_{1,2}$,  at least one of them has a positive real part and the other a negative real part, which signals a saddle point and an unstable equilibrium of the first node.  On the other hand, both $\lambda_{3,4}$ have a negative real part, indicating the existence of a spiral and a stable equilibrium of second and higher modes\cite{Slemrod}. The solution of unperturbed system~\cite{Holmes1979,Slemrod}, which is known to exist in the present case for the Hamiltonian given in eqn.(\ref{Ham}) is:
\begin{equation}
z_{0}(t)=\begin{pmatrix} C_{1}\sech(at)\\0\\C_{2}\sech(at)\tanh(at)\\0
\end{pmatrix}
\label{homo},
\end{equation}
where 
\begin{equation}
a=\left(\psi-Aq^{2}\right)^{\frac{1}{2}}\,,  \quad C_{1}=\frac{av_{0}^{3}}{2Q}\sqrt{\frac{3}{5\pi}}~~\text{and}~~C_{2}=-a C_1 \, .
%=[-\frac{4 \pi Q^2}{3 v_{0}^4} + \frac{9 \pi}{128} (-6 k + 18\pi T_{0} v_{0})-Aq^{2}]^{\frac{1}{2}}
\label{aeqn}
\end{equation}
Having established the presence of a homoclinic orbit in eqn. (\ref{homo}) connecting the origin to itself in the unperturbed system, we now add the small temporal perturbation, and compute the
Melnikov function defined earlier in eqn. (\ref{melnikov}) to  be:
\begin{equation}
M(t_{0})=-\int_{-\infty}^{+\infty}\Big[A_{1}\gamma\cos(\omega t)\chi\xi+A_{2}\gamma\cos(\omega t)\xi^{3}\chi-q\mu_{0}A_{3}\xi^{2}\chi^{2}\Big]\, ,
\end{equation}
where $\chi=\sech(a(t-t_{0}))$~ and~$\xi=\tanh(a(t-t_{0}))$. Further,
$A_{1}=(\frac{9\pi^{2}k\alpha}{4}+\frac{81\pi^{2} v_{0}^{2}}{128})C_{2}, A_{2}=\frac{243\pi^{2}C_{2}C_{1}^{2}}{512}$ and $A_{3}=C_{2}^{2}$. The evaluation of $M(t_{0})$ is best done using the residue theorem, resulting in:
\begin{eqnarray}
&& M(t_{0})
%=(A_{4}\pi \sech(\frac{\pi\omega}{2a}))\gamma\omega\sin(\omega t_{0})-q\mu_{0}A_{3}\frac{\pi}{2a}\nonumber\\
=\textbf{N}\gamma\omega\sin(\omega t_{0})-q\mu_{0}\textbf{I}\, ,
\end{eqnarray} 
where 
\begin{eqnarray}
\textbf{N}=A_{4}\pi \sech\left(\frac{\pi\omega}{2a}\right)~~ \text{and}~~\textbf{I}=\frac{\pi A_{3}}{2a}\, ,
\end{eqnarray}
with
\begin{eqnarray}
A_{4}=\frac{C_{2}(\frac{81}{128}\pi^{2}v_{0}^{2}+\frac{9}{4}\pi^{2}k\alpha)}{a^{2}}+\frac{C_{1}^{2}C_{2}(\omega^{2}+a^{2})}{16a^{4}}\frac{81\pi^{2}}{64},~~\text{and}~~
A_{3}=C_{2}^{2}\, .
\end{eqnarray}
$M(t_{0})$ has simple zeros at $\textbf{N}\gamma\omega\sin(\omega t_{0})-q\mu_{0}\textbf{I}=0$, giving the bound 
\begin{equation}
\left|\frac{q\mu_{0}\textbf{I}}{ \textbf{N}\gamma\omega}\right|\leq 1 \, .
\label{eqn36}
\end{equation}
%Since we can deduce that sufficiently small perturbation $\epsilon> 0$ can generate chaotic behavior due temporal thermal fluctuation.
Further, eqn.(\ref{eqn36}) translates in to a critical value for the perturbation parameter $\gamma$ of eqn. (\ref{pert}), as follows:
\begin{eqnarray} \label{critical}
\gamma_{{\rm critical}} = \left(\frac{\sqrt{3}512a^{5}v_{0}^{3} q\cosh\left(\frac{\pi\omega}{2a}\right) C_{1}\mu_{0}}{18\sqrt{5} Q\pi^{3/2}\omega \left(256a^{2}k\alpha+9a^{2}C_{1}^{2}+9\omega^{2}C_{1}^{2}+72a^{2}v_{0}^{2}\right)}\right)\, .
\end{eqnarray} 
One notes from eqn. (\ref{critical}) that, a small perturbation with $\gamma > \gamma_{{\rm critical}}$ guarantees the transversal intersection of stable and unstable manifolds, including the possible occurrence of Smale horseshoe chaotic motion~\cite{Holmes1990,Smale}. Chaotic behavior can be noted from figure-(\ref{fig:gammacritical1}), where a numerical plot of time evolution of equations of motion in (\ref{eom1})-(\ref{eom4}) is presented (for simplicity, $x_2, u_2$ are set to zero). Plots in figure-(\ref{fig:gammacritical1})$(a)$ and figure-(\ref{fig:gammacritical1})$(b)$  show normal trajectories of the system, in the absence and presence of a small perturbation (but, for $\gamma < \gamma_{\rm critical}$), respectively. Figure-(\ref{fig:gammacritical1})$(c)$ shows the onset of chaotic trajectories for $\gamma > \gamma_{\rm critical}$. The value of $\gamma$ that needs to be chosen for chaotic behavior is shown as the shaded region in the figures-(\ref{fig:gammacritical2})$(a)$ and (\ref{fig:gammacritical2})$(b)$, which essentially correspond to the plots of  eqn. (\ref{critical}). Wherever not mentioned, all the parameters are taken to be unity and with out loss of generality, specific values of $\alpha$ and/or $Q$ etc., need to be chosen for obtaining the plots. In any case, the plots in figures-(\ref{fig:gammacritical2})$(a)$ and (\ref{fig:gammacritical2})$(b)$ are only suggestive and the actual expression of bound for $\gamma$ given in eqn. (\ref{critical}) is to be used for actual computation of the allowed range.
%%%%%%%%%%   
  %%%%%%%%%%%%%%%%%%%%%%%
%%%%%%%%%%%%%%%%%%%%
It is interesting to note from eqn. (\ref{homo}), that the homoclinic orbit does not exist for $Q=0$, as the non-linear term leading to such an orbit, is absent from the Hamiltonian in eqn. (\ref{Ham}). The non-linear term in the Hamiltonian in eqn.(\ref{Ham}) can be traced back to the $\bar{P}_{v,v,v}(v_{0},T_{0})$ term in eqn.(\ref{HamG}). As seen from eqn. (\ref{EffectiveEOS}), this term vanishes for $Q=0$.  Thus, we conclude that for neutral Gauss-Bonnet black holes, chaos under temporal perturbations does not occur, unless the black hole carries charge. Noting the importance of the non-vanishing nature of $\bar{P}_{v,v,v}(v_{0},T_{0})$, the above results can be generalized to more general black hole systems, by asking: what is the minimum power of $v$, that needs to be present in the equation of state for nonlinearity to appear in the Hamiltonian and lead to chaos? To answer this, let us assume  a relation such as $P\propto 1/v^n$, for a generic black hole in extended phase thermodynamics, where $n$ is the largest power of $v$ that occurs in a given black hole equation of state\footnote{This assumption is valid for most of the static black holes as the equation of state contains polynomials of $v$.} in a general dimension $d$. The condition to rule out non-linearity in the Hamiltonian in eqn.(\ref{HamG}) and absence of chaotic behavior is that $\bar{P}_{v,v,v}(v_{0},T_{0}) = 0$. Solving this equation, we get a relation between $n$ and $d$ as:
\bea\label{condition}
d=2, n>0, \quad d > i, n=d-i ~~{\rm for}~~i=3,4,5\, .
\eea 
Let us note that the conditions in eqn. (\ref{condition}) are obtained for any generic black hole in AdS with extended phase thermodynamic description and can be used to rule out chaos based on the  equation of state itself. For instance, as seen from eqn.(\ref{eqnstateGBQ}),  the largest power of $v$ in the equation of state for a neutral Gauss-Bonnet black hole in a general dimension $d$ is $n=4$. Either one of the conditions, given in eqn. (\ref{condition}), is always satisfied for any $d>4$ for $n=4$. We thus conclude that chaotic behavior under temporal perturbations would be absent for neutral Gauss-Bonnet black holes in any dimension. On the other hand, for charged GB black holes, there is a term in equation of state in  eqn.(\ref{eqnstateGBQ}), which contains higher powers of $v$ and it can be checked that the conditions in  eqn. (\ref{condition}) are not satisfied in any dimension. Thus, chaotic behavior is possible, once the perturbation parameters satisfy the constraints put forward in eqn. (\ref{critical}). These differences between neutral and charged black holes should be investigated further for better understanding, especially, because the spinodal region in the PV diagram continues to exist, irrespective of the presence of charge $Q$ or not. It is also important to note that many of above results depend crucially on the distinction made between thermodynamic volume $V$ and specific volume $v$ through the new effective equation of state in eqn. (\ref{EffectiveEOS}), while performing computations. \\

\noindent
We now extend the results obtained above to more general case of Lovelock black holes in higher dimensions and check for chaotic behavior.
  %%%%%%%%%%%%%%%%%%%%%%%
    \begin{figure}[h]
  	% \begin{wrapfigure}{l}{0.3\textwidth}
  	\begin{center}
  		{\centering
  		\subfloat[]{\includegraphics[width=2.2in]{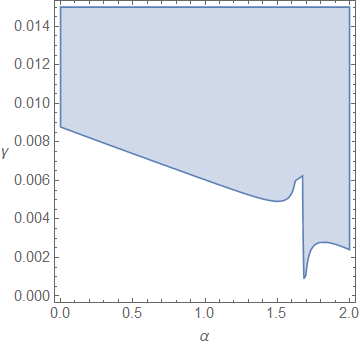} } \hspace {1.5cm}
  			\subfloat[]{\includegraphics[width=2.2in]{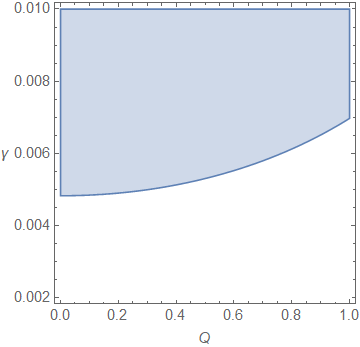} } \hspace {1.5cm}
  			\caption{Charged Lovelock Black holes  in $d=7$, with $v_0=1.44, T_0=0.1$: Plots of  (a) $\gamma$ vs $\alpha$ for $Q=2$ and (b) $\gamma$ vs $Q$ for $\alpha= 1$. Values of $\gamma$ in the shaded region lead to onset of chaotic motion }   \label{fig:gamma7d}
  		}
  	\end{center}
  	%\end{wrapfigure}
  \end{figure} 
The details of the action and PV critical behavior are discussed in detail in~\cite{Dehghani:2008qr,Mo:2014qsa} and we only need to recall the equation of state given as:
\begin{equation} \label{lovelockeqn}
P=\frac{T}{v}+\frac{32k\alpha T}{(d-2)^2v^3}+\frac{256k^2T\alpha ^2}{(d-2)^4v^5}-\frac{(d-3)k}{(d-2)\pi v^2}-\frac{16k^2(d-5)\alpha
}{(d-2)^3\pi v^4}-\frac{256k^3(d-7)\alpha ^2}{3(d-2)^5\pi v^6}+\frac{16^{d-3}(d-3)Q^2}{\pi(d-2)^{(2d-5)}v^{(2d-4)}} \, .
\end{equation} 
It is known that a Van der Waals type phase transition exists in these theories, together with a presence of spinodal unstable region. The procedure discussed in this section can be straightforwardly extended to the present case in all dimensions, starting from the equation of state in eqn.(\ref{lovelockeqn}) in general dimensions.
Applying the condition in eqn. (\ref{condition}) to the equation of state in eqn. (\ref{lovelockeqn}) above, chaos can be ruled out for neutral third order Lovelock black holes starting from dimension $d=7$. For the charged case, however, the nonlinear terms following from the relevant Hamiltonian in eqn. (\ref{HamG}) will be present and we see below that chaotic behaviour above a certain value of $\gamma$ persists.  We have computed analytically the expressions for Hamiltonian, Melnikov functions and associated bound on $\gamma$, but they are cumbersome and otherwise not very illuminating.  We suppress the expressions and present a plot of $\gamma $ vs $\alpha$ and $Q$ in seven dimensions in figure-(\ref{fig:gamma7d}), where the shaded parts show the allowed regions of $\gamma$ for which temporal chaos will be present. It is also useful to mention that the plots in figure-(\ref{fig:gamma7d}) are only suggestive and are obtained after choosing specific values of parameters being varied. For knowing the actual bound, the exact analytical expressions are to be used.\\

\noindent
Now, let us comment on the chaotic behaviour in the case of RN AdS black holes in general dimensions. In this case, the equation of state is given as~\cite{Belhaj:2012bg}:
	\begin{equation} \label{eqnsRNADS}
	P=\frac{T}{v}-\frac{d-3}{(d-2)\pi v^{2}}+\frac{(d-3)2^{4(d-3)}Q^{2}}{(d-2)^{2d-5}\pi v^{2(d-2)}}\, .
	\end{equation} 
%	For 5-Dimensional case thjis will take the following form
%	\begin{equation}
%	P=\frac{T}{v}-\frac{2}{3\pi v^{2}}+\frac{512Q^{2}}{243\pi v^{6}}
%	\end{equation}
It can be checked explicitly, that none of the conditions  in eqn. (\ref{condition}) are satisfied in any dimension and hence chaos will exist beyond a certain value of the perturbation parameter $\gamma$ in RN AdS black holes in any dimension. As the Gauss-Bonnet terms are total derivative terms in four dimensions, we can start comparing with the RN AdS case (by setting $\alpha=0$), starting from five dimensions.  In fact, setting $\alpha=0$ in eqn. (\ref{critical}), gives the limit on $\gamma$, beyond which chaos will exist in RN AdS black holes in five dimensions. The corresponding result in four dimensions was explicitly computed in~\cite{Chabab:2018lzf}. Thus, the conclusion that chaotic behaviour should be present in the four dimensional example of RN AdS black holes, studied in~\cite{Chabab:2018lzf}, is in conformity with our general condition in eqn. (\ref{condition}). 
%However, the bound on $\gamma$ has to be computed in each dimension seperately, as the corresponding values of thermodynamic quantities giving rise to the Maxwell equal area law vary significantly and depend on the number of dimensions. 
Furthermore, as $\alpha$ increases, it can be noted from figure-\ref{fig:gammacritical2}(a), that the GB system becomes chaotic for even smaller values of the perturbation parameter $\gamma$. \\
%Thus, the GB system is more sensitive in showing chaotic behaviour, for smaller temporal perturbations, as compared to the case of black holes in RN AdS background. \\
%Thus, one notes that the Gauss-Bonnet parameter $\alpha$ enhances the chaotic behaviour under temporal perturbations, as 

\noindent
To conclude this section, we note that the presence of charge $Q$ is necessary for triggering chaos under temporal perturbations in the extended thermodynamic phase space of black hole systems.
%%%%%%%%%%%%%%%%%%%%%%%%%%%%
\section{Spatial Perturbations and Chaos} \label{spac}
%%%%%%%%%%%%%%%%%%%%%%%%%%%%%

In this section, our aim is to study the effect of a small spatially periodic perturbation in the equilibrium state solutions about a sub-critical temperature given as follows~\cite{Slemrod}:
\begin{equation}
T=T_{0}+\epsilon\cos(qx) \, .
\label{temperature expression}
\end{equation}
Korteweig's theory gives the Piola stress tensor as~\cite{Slemrod}:
\begin{equation}
\tau=-P(v,T)-Av^{\prime\prime}
\label{eqn39}
\end{equation}
where $^{\prime}$ stands for $\frac{d}{dx}$. $P(v,T)$ is supplied by the GB black hole equation of state from eqn.(\ref{eqnstate}) and $T$ is absolute temperature with $A>0$. For zero body force balance of linear momentum, one sets $\tau^{\prime}=0$, giving $\tau=B=\text{constant}$. Thus, $B$ is the ambient pressure as $|x|\rightarrow\infty$; using this, eqn.(\ref{eqn39}) yields:
\begin{equation} \label{vel}
v^{\prime\prime}+P(v,T)=B \, .
\end{equation}
%We will concentrate on the heteroclinic and homoclinic orbits. 
Let us start by discussing the unperturbed system, where one starts by setting $T=T_{0}$ in eqn.(\ref{vel}). 
%\begin{equation}
%v^{\prime\prime}=B-P(v,T)
%\label{eqn41}
%\end{equation}
The fixed points of the system in eqn.(\ref{vel}) can be found, which are the specific volume corresponding to ambient pressure B for different given temperatures. We choose a set of sample temperatures, $0.8T_{c}$ and $0.7T_{C}$ and call the corresponding fixed points as $(v_{1},v_{2},v_{3})$ and  $(w_{1},w_{2},w_{3})$. 
For the case of $T_0= 0.8 T_c$, these are shown in figure {\ref{fig2}, with analogous construction assumed at 
 $T_0= 0.7 T_c$.
 \begin{figure}[h]
	\centering
	\includegraphics[width=7cm,height=5cm]{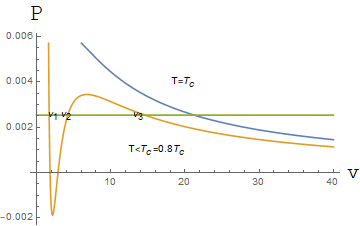}
	\caption{Charged GB Maxwell Equal Area construction for Q=1, $\alpha=1$, k=1, $T=0.8T_{c}$}
	\label{fig2}
\end{figure}
Let us note that Maxwell equal Area construction for Gauss Bonnet black holes done in~\cite{Belhaj:2014eha} is useful while plotting figure-{\ref{fig2}. Now, from eqn.({\ref{vel}}) and figure-{\ref{fig2}, one infers three different kinds of orbits in $v^{\prime}-v$ phase plane.
\begin{itemize}
	\item Case-1: In this case we choose the pressure in the range $P(v_{1},T_{0})<B<P(\beta,T_{0})$ and get a homoclinic orbit connecting a saddle point $v_{3}$ to itself . Corresponding phase orbits are shown in figures-{\ref{fig4}}$(a)$ and {\ref{fig4}}$(b)$ for charged and neutral black holes, respectively.
	\item Case-2: Choosing the pressure in the range $P(\alpha,T_{0})<B<P(v_{2},T_{0})$, results in a homoclinic orbit connecting a saddle point $v_{1}$ to itself, as in case-1 above. Corresponding phase orbits are presented in figures-{\ref{fig5}}$(a)$ and {\ref{fig5}}$(b)$ for charged and neutral black holes, respectively.
	\item Case-3: In this case the pressure is taken such that, $P(v_{1},T_{0})=B=P(v_{2},T_{0})$; This results in a heteroclinic orbit connecting $v_{1}$ with $v_{3}$. Corresponding phase orbits are shown in figures-{\ref{fig6}}$(a)$ and {\ref{fig6}}$(b)$ for charged and neutral black holes, respectively.
\end{itemize}
  %%%%%%%%%%%%%%%%%%%%%%%
    \begin{figure}[h]
  	% \begin{wrapfigure}{l}{0.3\textwidth}
  	\begin{center}
  		{\centering
  		\subfloat[]{\includegraphics[width=7cm,height=4cm]{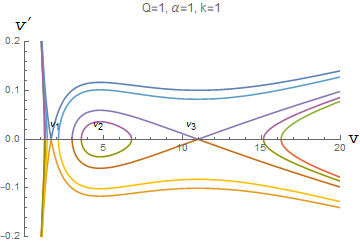} } \hspace {1.5cm}
  			\subfloat[]{\includegraphics[width=7cm,height=4cm]{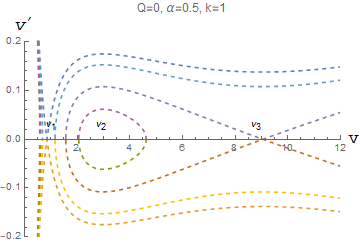} } 
  			\caption{Case1: (a) Charged Gauss-Bonnet with $v_{1}=1.68107,v_{2}=4.77519,v_{3}=10.9746$. (b) Neutral Gauss-Bonnet with $v_{1}=0.849379,v_{2}=2.97856,v_{3}=9.04772$ }  \label{fig4}
  		}
  	\end{center}
  	%\end{wrapfigure}
  \end{figure} 
%%%%%%%%%%%%%%%%%%%%%%%%%%%%%
%%%%%%%%%%%%%%%%%%%%%%%
    \begin{figure}[h]
  	% \begin{wrapfigure}{l}{0.3\textwidth}
  	\begin{center}
  		{\centering
  		\subfloat[]{\includegraphics[width=7cm,height=4cm]{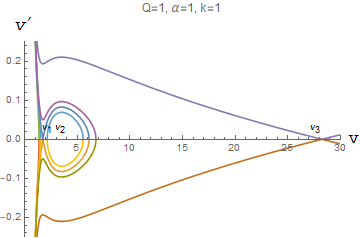} } \hspace {1.5cm}
  			\subfloat[]{\includegraphics[width=7cm,height=4cm]{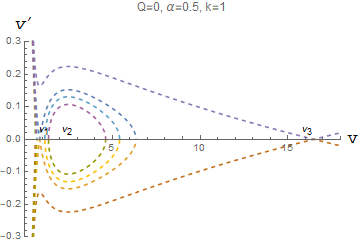} } 
  			\caption{Case2: (a) Charged Gauss-Bonnet with $v_{1}=1.73389,v_{2}=3.51509,v_{3}=28.2263$. (b) Neutral Gauss-Bonnet with $v_{1}=0.862346,v_{2}=2.47978,v_{3}=16.4746$. }  \label{fig5}
  		}
  	\end{center}
  	%\end{wrapfigure}
  \end{figure} 
%%%%%%%%%%%%%%%%%%%%%%%%%%%%%
 %%%%%%%%%%%%%%%%%%%%%%%
    \begin{figure}[h]
  	% \begin{wrapfigure}{l}{0.3\textwidth}
  	\begin{center}
  		{\centering
  		\subfloat[]{\includegraphics[width=7cm,height=4cm]{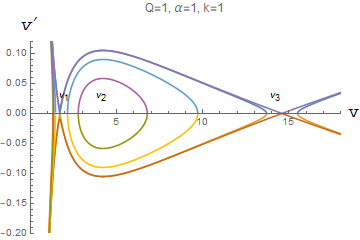} } \hspace {1.2cm}
  			\subfloat[]{\includegraphics[width=7cm,height=4cm]{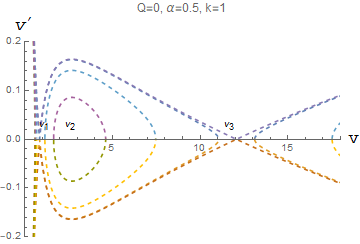} } 
  			\caption{Case3: (a) Charged Gauss-Bonnet with $v_{1}=1.69635,v_{2}=4.20704,v_{3}=14.593$. (b) Neutral Gauss-Bonnet with $v_{1}=0.855689,v_{2}=2.68773,v_{3}=12.0659$. }  \label{fig6}
  		}
  	\end{center}
  	%\end{wrapfigure}
  \end{figure} 
%%%%%%%%%%%%%%%%%%%%%%%%%%%%%
Including a small spatial perturbation given in eqn.(\ref{temperature expression}), we can rewrite the eqn.(\ref{vel}) for perturbed system as follows:
	\begin{equation}
v^{\prime\prime}=B-P(v,T_{0})-\frac{\epsilon\cos(qx)}{v}
\label{eqn42}
	\end{equation}
Melnikov function from eqn. (\ref{melnikov}) written suitably for spatially perturbed systems is:
	\begin{equation}
	M(x_{0})=\int_{-\infty }^{\infty}f(z(x-x_{0})){\Omega}_{n=1}g(z(x-x_{0}),x)dx
	\label{eqn43}
	\end{equation}
Setting $v^{\prime}=h$, eqn.(\ref{eqn42}) converts to a set of first order equations as:
	\begin{eqnarray}
	&&v^{\prime}=h\nonumber\\
	&&h^{\prime}=B-P(v,T_{0})-\frac{\epsilon\cos(qx)}{v}
	\end{eqnarray}
As in the previous section, writing general solutions for (homoclinic or heteroclinic) orbit as:
	\begin{equation}
	z(x)=\begin{pmatrix} 
	v_{0}(x-x_{0})\\h_{0}(x-x_{0}) \, ,
	\end{pmatrix}
	\label{homoclinic orbit for spatial chaos},
	\end{equation}
and using them in eqn.(\ref{eqn42}), one can write
	$f(z(x-X_{0}))$ and $g(z(x-x_{0}),x)$ as
	\begin{equation}
		f(z(x-X_{0}))=\begin{pmatrix} h_{0}(x-x_{0})\\B-P(v_{0}(x-x_{0}),T_{0})\,
			\end{pmatrix}\, , \quad
		g(z(x-X_{0}),x)=\begin{pmatrix} 0\\-\frac{\cos(q,x)}{v_{0}(x-x_{0})}\, 
	\end{pmatrix}
	\, .
	\end{equation}
Using these in eqn.(\ref{eqn43}), Melnikov function is finally:
$$M(x_{0})=-\int_{-\infty }^{+\infty}\frac{h_{0}(x-x_{0})\cos(qx)}{v_{0}(x-x_{0})}dx$$ 
Changing variables to $R=x-x_{0}$ , the Melnikov function becomes:
$$M(x_{0})=-L\cos(qx_{0})+W\sin(qx_{0})$$
with
\bea
&& L=\int_{-\infty }^{\infty}\frac{h_{0}(R)\cos(qR)}{v_{0}(R)}dR \, , \quad
W=\int_{-\infty }^{\infty}\frac{h_{0}(R)\sin(qR)}{v_{0}(R)}dR \, .
\eea
From the structure of Melnikov function and following the arguments in~\cite{Slemrod}, $ M(x_{0})$ always possesses simple zeros, signalling chaos. The results of this section can be carried over to charged Lovelock black holes in various dimensions and we have checked that the general features found in this section continue to exist, namely the system exhibits homoclinic orbits for the cases-1 and 2 discussed above; For case-3, the system has homoclinic as well as heteroclinic orbits. The presence of chaos under spatial perturbations is found in all three cases in Lovelock black holes in higher dimensions, irrespective of whether the charge is present or not, unlike the case of temporal perturbations discussed in last section.

%%%%%%%%%%%%%%%%%%%%
\section{Conclusions} \label{conclusions}
%%%%%%%%%%%%%%%%%%%%
In this work, we studied the emergence of chaotic behaviour under temporal and spatial perturbations in the spinodal region of charged and neutral Gauss-Bonnet black holes in extended thermodynamic phase space.  The perturbed Hamiltonian system corresponding to the motion of the fluid in the spinodal region, following from the black hole equation of state was obtained and shown to possess nonlinear terms giving homoclinic/heteroclinic orbits in phase space. Analysis of the zeroes of the appropriate Melnikov functions gives information about the onset of chaos in the thermodynamic phase space. As regards temporal perturbations, chaotic behaviour is found to be present in charged GB black holes in five dimensions. The zeros of the Melnikov function give a bound on the perturbation parameter for chaos to exist. This was computed analytically, such as the one in eqn.(\ref{critical}) and depends on the charge $Q$ and the GB coupling $\alpha$.  It is important to note that in this paper, the computations were performed explicitly in five dimensions, as the Gauss-Bonnet term is a total derivative in four dimensions and does not effect the black hole solution. Setting  the GB coupling $\alpha=0$ in eqn. (\ref{critical}), gives us the bound on $\gamma$ for chaos to exist, for the corresponding RN AdS black holes in five dimensions. These general conclusions hold true in any dimension, based on the condition in eqn. (\ref{condition}). For instance, based on eqns. (\ref{condition}) and (\ref{eqnsRNADS}), it can be said that in four dimensions, chaotic behaviour should exist beyond a certain value of $\gamma$. This result is in conformity with the analysis in~\cite{Chabab:2018lzf} where the expicit value of $\gamma$ was computed in four dimensions. When $\alpha$ is non-zero, we find that the onset of chaotic behaviour occurs for even lower values of the temporal perturbation parameter $\gamma$, showing the sensitivity of chaotic behaviour in the presence of GB terms. Intriguingly, the chaotic behaviour under temporal perturbations is not present for neutral GB and Lovelock black holes in general dimensions, which needs to be investigated further.  A general condition was derived in eqn.(\ref{condition}), which can be used to rule out chaos under temporal perturbations in general dimensions by analysing the equation of state provided by the black hole.  In every dimension, for chaos to exist, the highest power of $v$ in the equation of state,  cannot be lower than a certain value (as governed by eqn. (\ref{condition})).  The reason for existence of chaotic behaviour in charged black holes is that the term dependent on charge $Q$ in the black hole equation of state, increases qudratically with $v$, as the number of dimensions increases. This can be seen from the last term in the GB and RN AdS equations of state given in eqn. (\ref{eqnstateGBQ}) and (\ref{eqnsRNADS}), respectively. Thus, chaotic behaviour crucially depends on the number of dimensions as well as the equation of state of the black hole.  It is also important to mention that many of above results depend crucially on the distinction made between thermodynamic volume $V$ and specific volume $v$. In particular, we expressed the pressure $P$ in terms of the specific volume $v$,  through the new effective equation of state in eqn. (\ref{EffectiveEOS}), while performing computations. The use of effective equaion of state in eqn. (\ref{EffectiveEOS}) is important in concluding the existence or non-existence of chaos\\

\noindent
Under spatial perturbations, existence of homoclinic and heteroclinic orbits is found to exist in charged as well as neutral GB black holes and the phase space plots were given. The extension of results to Lovelock black holes in higher dimensions was discussed. It would be interesting to understand the holographic aspects of these chaotic behaviour, particularly found in the unstable small/large black hole phase transition domain, not just in GB black holes, but also in other charged and neutral black holes in AdS, considering other stringy corrections. More importantly, the absensce of chaos in neutral black holes needs to be understood better.

\section*{Acknowledgements}
The authors would like to thank the anonymous referee for helpful suggestions.

%\bibliography{mybibfile}

\end{document}